\documentclass[aps,twocolumn,aps,showpacs]{revtex4}

\usepackage[dvips]{graphicx}

\begin{document}

\title{Antilocalization in Coherent Backscattering of Light in a
Multi-Resonance Atomic System}
\author{D.V. Kupriyanov and I.M. Sokolov}
\affiliation{Department of Theoretical Physics, State Polytechnic
University, 195251, St.-Petersburg, Russia}
\email{Kupr@quark.stu.neva.ru}

\author{M.D. Havey}
\affiliation{Department of \ Physics, Old Dominion University, Norfolk, VA 23529}
\date{\today }
\email{mhavey@physics.odu.edu}

\begin{abstract}
Theoretical prediction of antilocalization of light in ultracold
atomic gas samples, in the weak localization regime, is reported.
Calculations and Monte-Carlo simulations show that, for selected
spectral ranges in the vicinity of atomic $^{85}$Rb hyperfine
transitions, quantum coherence in optical transitions through
nondegenerate hyperfine levels in multiple light scattering
generates destructive interference in otherwise reciprocal
scattering paths. This effect leads to enhancement factors less
than unity in a coherent backscattering geometry, and suggests the
possibility of enhanced diffusion of light in ultracold atomic
vapors.
\end{abstract}

\maketitle

%% 34.50.Rk, 34.80.Qb - Laser modified scattering and reactions
%% 42.50.Ct - Quantum description of interaction of light and
%%            matter, related experiments

\section{Introduction}
An area of developing interest in fundamental atomic physics
research is the influence of disorder on atomic or on mesoscopic
dynamics.  As an example, recent theoretical results \cite
{Damski} have indicated that disorder-induced transitions from a
Bose-Einstein Condensate to Bose or to Anderson glass phases is
possible, and within current experimental capability.  Recent
studies by Krug, et al. \cite {Krug} have shown how disorder
within the eigenenergy spectrum of a single atom can explain
anomalies in multiphoton microwave ionization cross-sections.
Another area of considerable current research interest is
localization of light, which may be described as a
disorder-induced phase transition in the transport properties of
electromagnetic radiation in strongly scattering random media.
Investigations in this area, originally stimulated by Anderson
\cite {Anderson} localization of electrons, have resulted in two
demonstrations of light localization in condensed matter systems
\cite {Wiersma1,Chabanov1}.

Recently, coherent multiple light scattering has been
experimentally observed in samples of ultracold $^{85}$Rb and Sr
atoms \cite {HaveyExp,FrenchExp}. Corresponding theoretical
results \cite {HaveyExp,KupriyanovTh,FrenchTh,Kupriyanov2} have
revealed fundamental mechanisms responsible for the observations,
and numerical simulations have had considerable success in
describing the experimental results. In all cases, the essential
physical mechanisms are due to interferences in multiple wave
scattering from the constituent atoms of the medium. For ultracold
atoms, and under otherwise not stringent conditions, the
interferences survive configuration averaging, thus generating
macroscopic observables. For radiation incident on a disordered
system, the effect manifests itself as a spatially narrow ($\sim
$1 mrad) cusp-shaped intensity enhancement in the nearly backwards
direction \cite{Sheng,LagTig}; this is referred to as the coherent
backscattering (CBS) cone. In the present case, the disorder is
principally due to the random locations of the atoms as found in a
typical ultracold atomic gas sample. The detailed angular profile
and magnitude of the enhancement depends on the polarization of
the incident and the detected light, and on the physical size of
the sample under study. For classical radiation scattering from a
$^{1}S_{0}\rightarrow $ $^{1}P_{1}$ atomic transition, the largest
possible interferometric enhancement is to increase the intensity
by a factor of two \cite{Sheng,FirstCBS,FrenchExp} over the
incoherently scattered background light.

Earlier \cite{HaveyExp} we have shown that energetically remote
atomic resonances can influence the spatial distribution of
backscattered intensity in the cone region.  For near-resonance
scattering, these transitions appear as a spectral asymmetry, in
certain polarization channels, in the interferometric enhancement.
In addition, for CBS from a completely oriented atomic gas, we
have found, in a double scattering limit, a near-resonance
enhancement approaching the classical limit of two
\cite{HaveyExp}. In the present paper, we present a surprising
theoretical result which is due to the presence of off-resonance
transitions.   We find, for non resonant scattering, but over a
quite wide spectral range, destructive interference in certain
polarization channels of the coherently backscattered light. This
results in so-called antilocalization \cite {Berkovits} in the
optical regime, with a persistent enhancement less than unity.
Such effects in electron transport have been widely studied in a
range of physical systems where spin-orbit interaction is
important \cite {Zumbuhl,Springer1}. Theoretical study of
polarization effects in coherent backscattering of s = 1/2 massive
particles by Gorodnichev, et al. \cite{gorodnichev} have shown
related novel effects

In the following sections we present a brief overview of the
physical system, including interferences generated by off
resonance light scattering from multilevel atoms. This is followed
by a summary of the essential theoretical ideas, and our approach
to simulate coherent multiple scattering in an ultracold atomic
gas. We then present theoretical results, in a double scattering
limit, illustrating the antilocalization effect and its physical
origin.

\section{Overview of the Physical System}

\subsection{Interference in nonresonant atomic light scattering}
The amplitude for quasielastic scattering of light which is
detuned several natural widths from any atomic resonance has
significant contributions from all energetically nearby atomic
transitions. The scattering cross-section is essentially given by
the Kramers-Heisenberg \cite {Louden} equation. To illustrate this
point, consider light scattering from a single ground level, and
two non degenerate excited levels.  As the dispersive part of the
scattering amplitude changes sign when tuning through a resonance,
there will be a minimum (near-zero) in the scattering amplitude in
the spectral region between the two levels \cite {Havey1, Havey2}.
The scattering amplitude changes sign again as the incident
radiation is tuned through this minimum. When we consider the
complexity of real atomic transitions, the spectral location of
the minimum depends on the polarization of the incident radiation,
and on the multipole distribution within the ground state of the
scattering atom. This behavior of the dispersion, when selected in
one of two otherwise reciprocal multiple scattering trajectories
is what is directly responsible for the existence of
antilocalization in atomic CBS.

\subsection{Brief overview of the theoretical treatment}
A theory of the CBS process in an ultracold atomic gas has been
developed recently by several groups \cite
{HaveyExp,KupriyanovTh,FrenchTh}. The theoretical development
essentially maintains earlier conceptions of weak localization in
the atomic scattering problem \cite{Shlyap}, and takes into
account the influence of the optical depth and sample size on the
character of the CBS cone. Our treatment \cite
{KupriyanovTh,HaveyExp} also accounts for spectral variations on
the CBS cone, and on the influence of far-off-resonance
transitions. More recently, we have generalized the theoretical
treatment to include effects associated with an
angular-momentum-polarized atomic medium as discussed in \cite
{CTLa,KupriyanovTh,Kupriyanov2}. In both our earlier theoretical
approach \cite{KupriyanovTh}, and in the present report, the
general analytical development was realized by a Monte-Carlo
simulation of coherent multiple scattering in an ultracold ($T<50$
$\mu K$), Gaussian-shaped sample of gaseous $^{85}$Rb atoms
confined to a magneto optical trap.  These simulations, which
included experimental parameters to characterize the sample, were
in excellent agreement with our experimental results \cite
{HaveyExp}.

The most important modification of the simulation procedure,
comparing with that which we made earlier, is inclusion of
mesoscopically averaged polarization effects for light propagating
in the atomic sample. The problem of light propagation through a
sample consisting of atoms with arbitrary polarization in their
angular momentum was discussed first by Cohen-Tannoudji and
Lalo\"{e} \cite{CTLa}. In those papers, such well known effects of
optical anisotropy of an atomic vapor as birefringence, gyrotropy
and dichroism were connected with the formalism of the irreducible
tensor components of atomic polarization, i.e. with the
orientation vector and the alignment tensor. In the Green's
function formalism, the problem of light propagation through the
polarized atomic vapor was later discussed in Ref.
\cite{Kupriyanov1}. The technique developed there allows one to
obtain an analytical solution for the Green's function in many
practically important cases. Basically, the retarded-type Green's
function can be found as a solution of a certain type of Dyson
equation, which reveals the normal vacuum wave equation modified
by the contribution of the polarization operators on the
right-hand side. These operators include all the polarization
sensitive effects for forward propagating light. In turn, that
lets us make a self-consistent Monte-Carlo simulation of the CBS
process, including all the important polarization-sensitive
effects. For more details we address the reader to
\cite{Kupriyanov2}.

Finally we emphasize that in all cases the simulations are made
for conditions quite close to those in the experiment. These
conditions include sample size, temperature, shape and density,
and the characteristic intensity of the CBS laser beam. \ These
conditions are such that recurrent scattering may be neglected,
and such that saturation of the atomic transition is also
negligible. In simulations of thermal effects, and of the
influence of atomic magnetization on the coherent backscattering
enhancement, more severe conditions are used in order to
illustrate the possible range of influence of these effects.

\section{Results}

\subsection{Spectral variation of the CBS enhancement and antilocalization}

In this communication, we consider the helicity of the incident
radiation to be antiparallel to the orientation of the atomic
vapor. This geometry requires special preparation since, because
of optical pumping, there is a tendency to reorient the collective
spin vector of the atomic ensemble along the beam, especially
following a long interaction with the probe light, and after
accumulation of enough Raman-type transitions. However, at this
stage we can neglect the optical pumping mechanism and assume that
most of the atoms populate the $|F_0,m=-F_0\rangle$ Zeeman state.
In addition, we assume that the ensemble is located in an external
magnetic field directed along the light beam. Thus the photons
scattered via Raman channels are generally Zeeman frequency
shifted. We assume this splitting to be quite small, and
comparable or less than the natural line width of the atomic
transitions. However, because of typically very slow relaxation in
the ground state, such splitting is enough to consider any
Raman-type transition as an inelastic channel in the scattering
process. In the following section we will explain how such
quasi-elastic scattered modes can be resolved and selected in
experiment.

\begin{figure}[tp]
\includegraphics{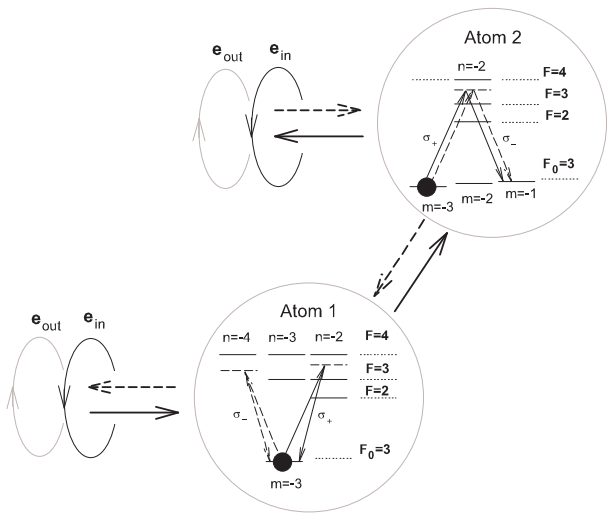}
\caption{Diagram explaining the CBS phenomenon in the helicity
preserving scattering channel for double scattering of circular
polarized light on an ensemble of ${}^{85}$Rb atoms oriented
opposite to the helicity vector of the probe beam. In this
example, double scattering is a combination of Rayleigh- and
Raman-type transitions. The solid and dashed lines indicate the
interfering direct and reciprocal scattering paths for probing
between $F_0=3\to F=4$ and $F_0=3\to F=3$ hyperfine transitions.}
\label{Fig.1}%
\end{figure}%

In Fig. \ref{Fig.1} we show double backscattering of a positive
helicity incoming photon on a system consisting of two ${}^{85}$Rb
atoms; the exit channel consists of detection of light also of
positive helicity. The two interfering channels, which are shown
here, repopulate atoms via Raman transitions from the
$|F_0,m=-F_0\rangle$ to the $|F_0,m=-F_0+2\rangle$ Zeeman
sublevel. In the direct path the scattering consists of a sequence
of Rayleigh-type scattering in the first step and of Raman-type
scattering in the second. In the reciprocal path, Raman-type
scattering occurs first, and the positive helicity photon
undergoes Rayleigh-type scattering in the second step. Since
identical helicities in incoming and outgoing channels have
opposite polarizations with respect to the laboratory frame, there
is an important difference in transition amplitudes associated
with Rayleigh process for these two interfering channels. Indeed,
in the direct path the $\sigma_+$ mode is coupled with $F_0=3\to
F=4$, $F_0=3\to F=3$ and $F_0=3\to F=2$ hyperfine transitions. But
in the reciprocal path, the $\sigma_-$ mode can be coupled only
with the $F_0=3\to F=4$ transition. As we see from the diagrams
shown in Fig. 1, where the probe light frequency $\omega_L$ is
scanned, for example, between the $F_0=3\to F=4$ and $F_0=3\to
F=3$ transitions, a unique spectral feature is found when the
scattering amplitudes connected the direct and reciprocal
scattering channels are equal in absolute value but have phases
shifted by an angle close to $\pi$. From an electrodynamic point
of view, such conditions are realized when, due to the asymmetry
in the Rayleigh-type transitions, the real part of the
susceptibility of the sample is positive for the $\sigma_-$ mode
and is negative for the $\sigma_+$ mode. Since, in a first
approximation, the amplitudes of the processes shown in Fig. 1
have opposite signs they will interfere destructively with {\it
anti-enhancement} of the light scattered in the backward direction
in the helicity preserving channel. Such an unusual behavior in
the CBS process is connected with the Raman nature of the helicity
preserving scattering channel.  We reiterate that, for this effect
to be observed, and not washed out by other competitive and
constructively interfering channels, both special light
polarization channels and selection of certain spectral domains,
are required.

In our example, as shown in Fig. 2, there are two competing
channels of double Raman-type scattering which can interfere
constructively. For equidistantly split Zeeman sublevels these
processes contribute in the helicity preserving channel of the
backscattering with the same frequency shift for the outgoing
photon as for the diagrams shown in Fig. \ref{Fig.1}. In an
experiment, possible selection of either destructively or
constructively interfering channels can be done by focussing
attention to their angular dependence with respect to the location
of the atomic scatterers. For an optically thin sample the
processes of Fig. 1 dominates for atoms located preferably along
the probe beam direction. The respective angular factor is
proportional to the probability to initiate the $\sigma_{+}$ or
$\sigma_{-}$ transition on the second atoms by the photon also
emitted on the $\sigma_{+}$ or $\sigma_{-}$ transition by the
first atom.

\begin{figure}[tp]
\includegraphics{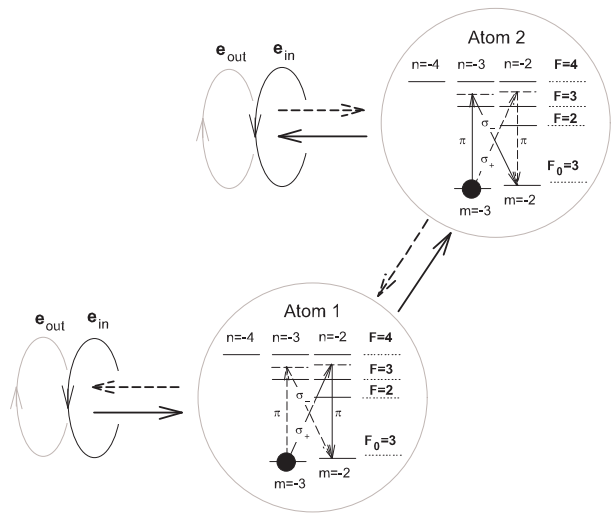}
\caption{Same as in Fig.\ref{Fig.1}, but the double scattering is
now a combination of successive Raman-type transitions.}
\label{Fig.2}%
\end{figure}%

This probability is given by
\begin{equation}
P_{++}(\theta)=P_{--}(\theta)\propto \frac{1}{4}(\cos^2\theta\,+\,1)^2.%
\label{4.4}%
\end{equation}%
In turn the processes shown in Fig. \ref{Fig.2} are dominant if
atoms are located preferably in the plane orthogonal to the probe
beam. The respective angular factor is proportional to the
probability to initiate the $\pi$ transition on the second atom by
the photon emitted on the $\pi$ transition by the first atom:
\begin{equation}
P_{\pi\pi}(\theta)\propto \sin^4\theta.%
\label{4.5}%
\end{equation}%
By comparing the angular distributions (\ref{4.4}) and
(\ref{4.5}), one can expect that, for a cigar-type atomic cloud,
stretched along the probe beam direction and squeezed in
orthogonal directions, the destructively interfering channels
should give the dominant contribution.

A more general way to select the destructively interfering
channels can be achieved if the Zeeman sublevels are
non-equidistantly split by applied fields. For heavy alkali atoms,
as in the current example of ${}^{85}$Rb, this can be done by
admixture of an external electric field. But it would be even more
straightforward to accomplish this for lighter alkali atoms, where
the splitting is always non-equidistant because of quadratic
Zeeman effect. In this case the contribution of the processes
shown in Fig. \ref{Fig.2} can be cancelled out and in double
scattering the destructive interference can drop the output
intensity down to near the zero level.

However, as can be verified, the effect of destructive
interference can be perfectly selected only for double scattering.
The destructive interference associated with the diagram of Fig. 1
can be constructive for more general diagrams, including higher
orders of scattering. We can expect that for a dense sample,
because of accumulation of multiple scattering contributions of
different orders, the effect of the destructive interference in
the backscattered light still survives but can be suppressed.
Realistic estimation of the remaining anti-enhancement factor can
be done only numerically.

\begin{figure}[tp]
\includegraphics{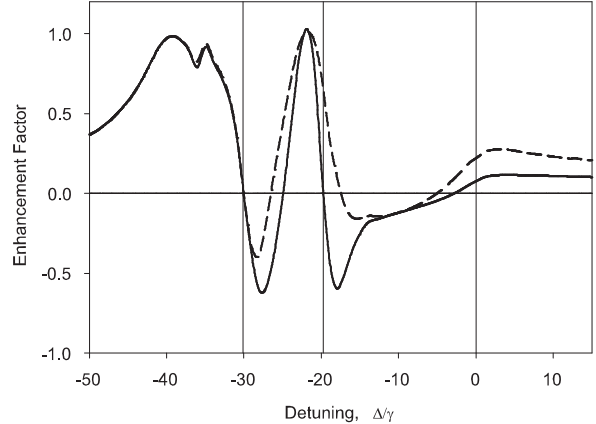}
\caption{The enhancement factor in the helicity preserving channel
for double scattering as a function of detuning
$\Delta=\omega_L-\omega_{43}$ for the probe laser scanning the
upper hyperfine manifold of ${}^{85}$Rb. The solid curve shows the
contribution of the processes depicted in Fig.\ref{Fig.1} and in
the dashed curve the constructively interfering channels shown in
Fig.\ref{Fig.2} were added. Vertical lines indicate the locations
of hyperfine resonances.}
\label{Fig.3}%
\end{figure}%

In Fig. \ref{Fig.3} we show the results of numerical simulations
of the enhancement factor in the case of pure double scattering
for a spherically symmetric dense Gaussian-type cloud and for the
Zeeman splitting in the ground state of $0.1\gamma$. In the graphs
plotted in Fig. 3 the optical depth has the same value of $b=1$
for each detuning laser frequency $\omega_L$ or corresponding
detuning $\Delta=\omega_L-\omega_{43}$. The solid curve shows the
contribution of the processes depicted in Fig. \ref{Fig.1}. The
frequency of the probe laser $\omega_L$ scans the whole upper
hyperfine manifold from the $F_0 = 3$ ground hyperfine level, and
there are two evident off-resonance spectral domains associated
with antilocalization. In these spectral ranges, the enhancement
factor is negative. It is interesting that even in such an ideal
situation, the anti-enhancement is less than 100\%. This is
because of the polarization dependence of the refractive index
along the light ray, which is characterized by the properties of
the Green propagation function; this essentially modifies the
results of the ideal conditions required for complete destructive
interference. The chain curve in Fig. \ref{Fig.3}. shows how the
enhancement factor would change if both the destructive and
constructive interfering channels shown in \ref{Fig.1} and
\ref{Fig.2} were taken into account. Comparison of these
dependencies suggests to us that antilocalization would be more
readily observed in a system with non-equidistant Zeeman
sublevels.

An essential and important point is that maximal constructive
interference in the CBS process is normally associated with time
reversed symmetry of the scattering amplitude. Indeed, the
reciprocal path is described by an amplitude which, in certain
cases, can be similar to a time reverse amplitude of the direct
path. That is the situation if there are only two interfering
channels and Rayleigh type scattering. But that is not the case
for processes appearing in the Raman channels and assisted by a
magnetic type interaction such as hyperfine interaction of nuclear
and electronic spins. The destructively interfering amplitudes
shown in Fig. 1 are not reversed in time. The complete time
symmetry needs that the sign of the field to be changed for the
time reverse amplitude, leading to a sign change in angular
momentum projectors. In our case the transition amplitude is in
fact determined by the matrix elements of only the operators
acting on electronic degrees of freedom and nuclear spin plays a
role similar to an external magnetic field with respect to such an
amplitude. In turn this means that the processes discussed for
isolated atoms will not be seen without internal magnetic
interactions us as a hyperfine interaction, which is in fact
obvious as far as the ground state of atom would be a spin one
half state in this case.

Finally, we point out an interesting possibility suggested by our
results.  It is well known \cite {Sheng,LagTig} that constructive
interference in recurrent scattering can reduce the effective
diffusion constant in a randomly scattering medium. Alternately,
this means that optical diffusive transport through the medium
suffers from an enhanced resistance to the diffusive current.  In
direct analogy with our results, destructive interferences in
recurrent scattering should lead to modified diffusion of light in
the medium.  Although the regime of recurrent scattering has not
yet been reached in experiments on ultracold atomic gases, it is
well within the current capabilities of experimental ultracold
atomic physics.  A useful observable in that case would be the
analogue of the recent experiments of Labeyrie, et al. \cite
{Labeyrie4}, where the time dependence of light emerging from an
ultracold cloud of $^{85}$Rb was measured. The rate of emergence
of diffusive light should be enhanced in the antilocalization
spectral regime.

\subsection{Proposed experimental investigation using light-beating spectroscopy}
In the previous section it was indicated that the calculations
were done in a small static magnetic field on the order of a few
gauss. This produces a slight ground state Zeeman shift which may
be exploited to select different quasi-elastic Raman channels
contributing to the multiple scattering process.  The essential
tool is the light beating method, which is a common technique of
high resolution spectroscopy. In an application of this method to
observation of antilocalization, the scattered light should be
mixed with a local oscillator in a balanced heterodyne detector.
This coherent and high quality stable mode could be a portion of
the monochromatic probe light. Then any Raman scattering channel
can be observed in the resultant photocurrent spectrum because of
the relatively low frequency of the beat note associated with the
Zeeman shift compared with the local oscillator.  With this
approach, the Raman scattering could be observed as a resonant
feature in the photocurrent spectrum at the reference frequency
equal to the Zeeman splitting between sublevels
$|F_0,m=-F_0+2\rangle$ and $|F_0,m=-F_0\rangle$.

Here we mention only that the photocurrent spectrum reproduces the
spectral profile of the scattered light emerging the sample and
responding to monochromatic incident light. While probing the
sample in an off-resonant domain, the spectral distribution is
described by the effect of any Raman shift and also by broadening
of the originally monochromatic coherent wave as a result of
atomic motion. For the sake of simplicity, let us consider an
optically thin atomic sample (in the frequency range associated
with antilocalization) and ignore effects associated with optical
depth and anisotropy; these features may be quite important under
real experimental conditions. Then the contribution of single
scattering is described by the following spectral profile
\begin{equation}
I_1(\omega)\propto \int\!\! d\tau\, e^{i(\omega-\omega_R)\tau}\,%
\left\langle\exp\left[2i\frac{\omega_L}{c}(z(\tau)-z(0))\right]\right\rangle%
\label{4.6}%
\end{equation}%
where $\omega_L$ is the probe light laser frequency and $\omega_R$
is the carrier frequency for light scattered via the Raman
channel. The angle brackets denote averaging over the velocity
distribution for any atom randomly moving in space. For steady
state conditions, the stochastic function $z(\tau)=z(0)+v_z\tau$,
where $v_z$ is a random atomic velocity projected on the
$z$-direction, and shows the atomic displacement along the
$z$-direction for a short time increment $\tau$. The spectral
profile for double scattering can be expressed as
\begin{equation}
I_2(\omega)\propto \int\!\! d\tau\, e^{i(\omega-\omega_R)\tau}\,%
\left\langle\exp\left[i\frac{\omega_L}{c}(s_{12}(\tau)-s_{12}(0))\right]\right\rangle%
\label{4.7}%
\end{equation}
where $s_{12}(\tau)$ denotes the displacement of the scattering
loop for any pair of atomic scatterers. This displacement is
associated with the random motion of each atom, and includes all
the spatial coordinates with no preference for any particular
$z$-direction. The averaging in (\ref{4.7}) is extended over the
velocity distribution of two atoms. In this expression we ignore
Sagnac-type retardation effects for reciprocal paths; we are
neglecting the small displacement of atoms during the time of
light propagation between the scatterers.

From these equations, we see that the spectral profiles of single
and double scattering are similar but not identical. Thus it would
be possible to resolve these terms in a sufficiently precise
spectral analysis. The most critical experimental aspect would be
to resolve the Raman components in comparison with the Doppler
broadening, which should then be much less than Zeeman splitting.
This means that the atoms should be quite cold and heating effects
associated with the probe light interaction would be undesirable.
Such conditions are readily achievable using techniques of
ultracold atomic physics; similar conditions applied to our
previous measurements of coherent backscattering from ultracold
atomic Rb \cite {HaveyExp,KupriyanovTh} As a preliminary estimate,
we compared the total intensities given by the integrals in
(\ref{4.6}) and (\ref{4.7}) and of similar expressions for higher
orders of multiple scattering over $\omega$. This gives us the
enhancement factor for such a Raman-type scattering channel, which
could be expressed in terms of cross sections, as
\begin{equation}
X_{EF}=\frac{d\sigma^{S}\,+\,d\sigma^{L}_{\Sigma}\,+\,d\sigma^{I}_{\Sigma}}%
{d\sigma^{S}\,+\,d\sigma^{L}_{\Sigma}}%
\label{4.8}%
\end{equation}%
where the denominator contains the contributions of single
scattering and ladder terms of all scattering orders and the
numerator has an additional interference term. For a sample where
the spectrally-dependent optical depth $b$ in the range of the
enhancement minimum is varied near unity, the enhancement factor
changes in the interval of a few percent and has a rather stable
minimum value near $-0.03$. An effect of such size is readily
within current experimental capability. The relatively small value
can be explained by the fact that double scattering cannot fully
compete with the contributions from single scattering. As we have
verified by numerical simulations, in a dense sample with $b$
increased to ten, and including higher orders of scattering, the
effect of destructive interference appearing in double scattering
also survives in the level of a few percent, but is partially
washed out by the losses of light via other scattering channels
and by effects of optical anisotropy.

\section{Summary}
A theoretical study of spectral variations in the coherent
backscattering enhancement factor, for a multi-resonance atomic
gas, has been reported. Numerical simulations of the variations,
which consider the influence of hyperfine interferences in
multiple light scattering, and atomic magnetization, show that
antilocalization, where there is destructive interference in the
coherent backscattering cone, can occur in the optical regime.
This, in turn, suggests the possibility of enhanced diffusion of
light in an ultracold atomic gas. Finally, an experimental
approach to observation of the predicted effects, using
light-beating spectroscopy, is proposed.

\begin{acknowledgments}
Financial support for this research was provided by the National
Science Foundation (NSF-PHY-0099587, NSF-INT-0233292,
NSF-PHY-0355024), by the North Atlantic Treaty Organization
(PST-CLG-978468), and by INTAS (INFO 00-479). D.V.K. would like to
acknowledge financial support from the Delzell Foundation, Inc.
\end{acknowledgments}

\end{document}